# Structure and kinematical properties of the Galaxy at intermediate galactic latitudes

D.K.Ojha[1], O.Bienaymé[2], A.C.Robin[2,3], M.Crézé[2], and V.Mohan[4]

[1] Inter-University Centre for Astronomy and Astrophysics, Post Bag 4, Ganeshkhind, Pune 411 007, India
[2] Observatoire de Strasbourg, CNRS URA 1280, 11 rue de l'Université, F-67000 Strasbourg, France
[3] Observatoire de Besançon, 41 bis, Av. de l'Observatoire, BP 1615, F-25010 Besançon Cedex, France
[4] U. P. State Observatory, Manora Peak 263 129, Nainital, India

**Abstract.** We have carried out a sample survey in UBVR photometry and proper motions in different directions in the Galaxy, as part of an investigation of galactic structure and evolution. Three fields in the direction of galactic anticentre ($l = 167°$, $b = 47°$), galactic centre ($l = 3°$, $b = 47°$) and galactic antirotation ($l = 278°$, $b = 47°$) have been surveyed. Using our photographic photometry, we determine photometric distances for a sub-sample of stars in the color range $0.3 \leq (B-V) \leq 0.9$. The stellar space velocities (U, V and W) are derived directly from the measured proper motions and distances.

Using our new data together with wide-area surveys in other fields available to date, we discuss the radial and vertical structure of the Galaxy. We have derived the density laws for stars as a function of distance from the galactic plane for each absolute magnitude interval. The density laws for stars with $M_V \geq 3.5$ follow a sum of two exponentials with scale heights of 260±50 pc (thin disk) and 760±50 pc, respectively. This second exponential corresponds to a thick disk with a local density of $7.4^{+2.5}_{-1.5}$ % relative to the thin disk. The scale lengths for these two populations are respectively 2.3±0.6 kpc and 3±1 kpc.

The kinematical distribution of F and G–type stars have been probed to distances up to 3.5 kpc above the galactic plane. A new value for the solar motion has been determined from moderately distant stars ($1 < z < 2$ kpc). It is consistent with local determinations and implies that there is no large motion of the LSR relative to the mean motion of stars at 1-2 kpc above the galactic plane. The rotational velocity curve is found flat in the solar neighborhood. The radial gradient in velocity dispersions has been determined for the thin disk population. The thick disk appears as a kinematically distinct population from the thin disk and shows no vertical gradient. A multivariate discriminant analysis is also used to distinguish the thick disk from the thin disk and to estimate its asymmetric drift. It is found to be 53±10 km/s, independent of the galactic radius. Of the many models that have been proposed for the origin of the thick disk, the evidence at present seems to favour a model in which thick disk formed through the rapid dynamical heating of an early disk by satellite accretion into the disk.

## 1. Introduction

The concept of stellar populations has played a critical role in the development of modern studies of stellar evolution, galactic structure and galactic evolution. To understand fully the stellar populations and their differences would take us a long way toward a complete theory of galactic structure and evolution. Yet, there are many difficult questions that have been very unyielding to final solutions, such as, how can we define the populations in a way that is physically meaningful, how many discrete populations really are there, and how can we unambiguously separate the stellar constituents of the populations for detailed analyses ?

It was only with the discovery of the asymmetric drift of high-velocity stars (Strömberg 1924; Oort 1922, 1926), it became clear that multiple overlapping components would be needed if an ellipsoidal model were to provide an adequate representation of the data. After the identification of the galactic centre (Shapley 1918), and the discovery of the differential rotation of the galactic disk (Oort 1922, 1927; Lindblad 1925, 1927, 1959; Strömberg 1924), the multicomponent concept was taken to the limit with Lindblad's 1927 (see also 1959) proposed division of the Galaxy "into an infinite number of 'sub-systems' of varying mean speed of rotation". Strömberg's (1924) comprehensive analysis of the available data on stellar distances, proper motions, and radial velocities in the mid-1920s was among the first to show the differences in the velocity ellipsoids of various stellar groups. His kinematic separation into groups was the beginning of the classification of the galactic components that eventually led to the population concept.

The pioneering work on populations was carried out by Baade during the 1940's and 1950's using mostly observations of galaxies in the Local Group. In 1944, Baade introduced the idea of two basic stellar populations (Baade 1944, 1958a). Unlike the kinematic components discussed previously by Strömberg (1924) and Oort (1922, 1926), Baade's two populations were defined according to the morphology of the respective color-magnitude diagrams. Population I stars are stars with photometric characteristics similar to those found in open clusters and stellar associations (e.g. luminous early-type stars, classical Cepheids), and Population II stars are similar to the stars found in globular clusters (e.g. RR Lyrae stars, subdwarfs, Population II Cepheids).

At the conclusion of the 1957 Vatican conference on stellar populations, five stellar populations were advocated : extreme population I, older population I, disk population, intermediate population II, and halo population II. Excellent reviews on the subject have been written by King (1971), van den Bergh (1975), Mould (1982), Sandage (1986) and others.

Throughout the last decade there has been further debate over the structure and history of our Galaxy. One of the main controversies concerns the existence of a thick disk, i.e., a component with characteristics intermediate to those of the thin disk and halo populations, and whether or not three components (rather than two, five or more components) are needed to account for star counts, proper motion distributions and other observations. There are now strong evidences that our Galaxy is not simply described by the traditional disk and a halo, but that there must exist a third component, a thick disk, which is dominated over distances on the order of kiloparsecs above the plane, and intermediate in kinematics and metallicity between the disk and halo populations. The idea of "third component" has apparently been around in one form or another for decades, at least since the time of the 1957 Vatican conference (see Blaauw 1965). The concept of a thick disk came into existence after Gilmore & Reid (1983) showed need for it through their inability to fit two-component models to starcount data. Since Gilmore & Reid (1983), other starcount surveys have reached similar conclusions (Yoshii et al. 1987; Sandage 1987; Fenkart 1988). Proponents of a three-component model have raised questions about the nature of the thick disk, in particular, whether it is kinematically discrete (Wyse & Gilmore 1986; Sandage & Fouts 1987; Gilmore, Wyse & Kuijken 1989; Soubiran 1993ab) and/or chemically discrete (Carney et al. 1989; Carney, Latham & Laird 1990). Majewski (1992) have shown that an intermediate component must exist in order to account for a bimodal distribution in both the V-velocity and UV-excess up to 5.5 kpc above the galactic plane, far beyond the region dominated by the traditional thin disk of scale height $\sim$ 260 pc or less. The population of "disk" globular clusters and the apparent excess of intermediate - metallicity stars with vertical scale height $h_z \sim$ 1 kpc, represent also the evidence for a thick disk. (Baade 1958b; Morgan 1959; Zinn 1985; Armandroff & Zinn 1988).

It should be stressed here that starcount analysis alone do not warrant the reality of thick disk stars as a third major component of the Galaxy, because adding new component(s) in models always gives a superior fit to the data. In fact, current interpretations of thick disk stars have been ranging between a separate component (Gilmore & Wyse 1985; Bienaymé et al. 1990, 1992) and a high-velocity tail of the old disk (Norris 1987ab; see also Norris & Green 1989). This third discrete population could be the signature of a merger event early in the Galaxy's history occuring shortly after the disk had formed (Carney et al. 1989). In fact, the arguments regarding the existence of this third population bear on problems of galaxy formation and evolution. For example, "the G dwarf problem", that the local number of metal-poor long-lived stars is too small to be consistent with simple models of chemical evolution, may be solved straightforwardly by use of an intermediate-metallicity "reservoir" dispersed above and below the plane (Gilmore & Wyse 1986).

Studies of stellar populations often take as their starting point large-scale surveys, with selection criteria involving one or more observable variables. The definition of the survey has to be defined in order to ensure a good retrieval of the population parameters we look for. The physical location of the surveys is arguably the most important constraint. Obviously several survey locations need to be chosen in order to vary the mixing proportions of the various populations and to determine the scale heights, lengths and relative densities. On the other hand, the survey must probe to great distances so that the regions dominated by each of the Galaxy's populations are within the limits of the survey. For example it is necessary to reach V magnitude about 19 in order to reach domains dominated solely by the halo populations. Finally, in order to be confident that the data are free of systematic bias, it must be complete, i.e. all stars within a region of the sky must be surveyed without any selection criteria which bear on the problems being studied.

With these points in mind, we have carried out a sample survey in UBVR photometry and proper motions in 3 directions in the Galaxy. The 3 fields chosen are in the direction of galactic anticentre ($l = 167°$, $b = 47°$; Ojha et al. 1994a, 1995; hereafter GAC1,2), galactic centre ($l = 3°$, $b = 47°$; Ojha et al. 1994b; hereafter GC) and galactic an-tirotation ($l = 278°$, $b = 47°$; Ojha et al. 1995; hereafter GAR). In this paper, using our new data (GAC1,2, GC & GAR) together with similar data set towards the poles, we derive the kinematical and structural parameters of the thin disk and thick disk components of the Galaxy.

In the present study, we shall take the view that within a particular volume of the Galaxy, one deals with a finite mixture of discrete stellar populations. Our approach has been to see how far one can go using gaussian mixtures to model stellar populations. Thus, our discussion of mixture models will be limited to the following variables : the three kinematical variables (U, V & W), metal abundance [Fe/H] or distance. Under the assumption that within a stellar population, each of these variables follow a gaussian distribution, overall mixture distributions, and posterior mixing proportions are computed for two or three-component models. Parameters and error estimations are discussed in general terms. This (minimum) system of variables provides an ample basis for the discussion of issues that are relevant to the study of stellar populations. The purpose of the present work is :

(i) To present a statistical approach to the analysis of stellar populations through the application of finite mixture model (Stochastic-Estimation-Maximization method).

(ii) To examine the statistical properties of multivariate finite mixture models.

(iii) To estimate the parameters that describe the underlying stellar populations and for determining the number of discrete stellar populations.

The outline of the paper is as follows: In §2, we discuss briefly the use of multicolor photometry to derive the absolute magnitude and stellar distance for each star in our surveys. The derivation of the stellar space velocity UVW is explained in §3. In §4, we present the variation of the stellar number density as a function of distance from the galactic plane. From this, we deduce the structural parameters of the thin disk and thick disk populations. In §5, we derive the kinematics for the thin and thick disks from two statistical methods. In §6, we derive the components of the solar motion and discuss the consistency between the kinematics and density scales of the two populations (thin and thick disks) via the asymmetric drift relation. In §7, we discuss briefly our main results and the thick disk formation scenarios.

## 2. Stellar distance estimates

Photometric parallaxes were determined by estimating absolute stellar magnitudes, which were obtained from a $M_V$ versus B-V relation and taking into account the metallicity change as a function of the distance from the galactic plane. We made use of the BV photographic photometry to derive distances for F and G–type stars in our samples. The absolute magnitude-color relation for main-sequence stars of solar metallicity is derived as :

$$M_V = 5.83(B - V) + 0.753$$

This relation is valid on -0.1<B-V<1.5 color index range. A similar color-magnitude relation has been calculated for the Hyades main-sequence stars (Reid 1993). While there is good agreement between the two relations, the Hyades relation contains higher order of terms in B-V. Since the three fields (GAC1,2, GC & GAR) are at intermediate latitudes ($|b|=47°$), therefore we can neglect the influence of interstellar extinction in distance calculation.

### 2.1. Absolute magnitude correction

The color-absolute magnitude relation changes with metallicity so that metal-poor dwarf stars are fainter than metal-rich stars of the same temperature or color, and the opposite is true for giants. Consequently, due to the vertical metallicity gradient, the distance or absolute magnitude derived must be corrected according to the distance from the galactic plane.

The local value of the vertical metallicity gradient $\frac{\partial [Fe/H]}{\partial z}$ is not well determined. Current estimates cover most values from -0.225 $kpc^{-1}$ (Janes 1979) to -0.8 $kpc^{-1}$ (McClure & Crawford 1971; Jennens 1975; Yoss & Hartkopf 1979). Recent studies of field stars RR Lyrae (Butler et al. 1979, 1982) and globular cluster (Zinn 1980; da Costa, Ortolani & Mould 1982) abundances show little if any systematic trend in [Fe/H] within 1-2 kpc of the galactic plane, a dominance of objects with [Fe/H] $\sim$ -1.5 at 5 kpc, but little or no further systematic decrease beyond this distance.

This evidence for a smooth gradient from $\sim$ 1 to $\sim$ 5 kpc is complicated by the results of Rodgers, Harding & Sadler (1981) who derived solar abundances for main-sequence A stars up to $\sim$ 4.5 kpc from the plane, and of Yoss and Hartkopf (1979) who also derived near solar CN abundances for K giants up to 3 kpc.

In view of these uncertainties, Kuijken & Gilmore (1989) produced a model of the vertical metallicity gradient as indicative rather than conclusive estimates of the true gradient. Following Kuijken & Gilmore (1989) or Soubiran (1992), we assume the vertical gradient of metallicity is :

$$\frac{\partial [Fe/H]}{\partial z} = -0.3 \; kpc^{-1} \quad z < 5400 \; pc$$

$$[Fe/H] = -1.5 \quad z \geq 5400 \; pc$$

with $\quad [Fe/H] = +0.12 \quad$ for $\quad z = 0$.

The most recent study of the sensitivity of $M_V$ to [Fe/H] is by Laird et al. (1988). From a study of local subdwarfs with trigonometric parallaxes, they find the following relations between the metallicity-dependent UV-excess photometric parameter $\delta_{0.6}$ and the absolute magnitude difference $\Delta M_V$, relative to a Hyades main-sequence star of the same B-V color :

$$\Delta M_V = (\frac{2.31 - 1.04(B-V)}{1.594})(-0.6888\delta_{0.6} + 53.14\delta_{0.6}^2$$
$$- 97.004\delta_{0.6}^3) \quad 0 \leq \delta_{0.6} \leq 0.25$$

and

$$\Delta M_V = 0 \quad \delta_{0.6} < 0$$

The UV excess - metallicity relation was studied by Carney (1979), and can be approximated as :

$$\delta_{0.6} = -0.0776 + \sqrt{0.01191 - 0.05353[Fe/H]}$$

So we obtain a $\Delta M_V$ correction that depends on $z$. The absolute magnitudes calculated from the $M_V$ *versus* B-V relation was therefore corrected by these amounts. In practice, a new distance is calculated, a revised $\Delta M_V$ correction applied, and the procedure was repeated to convergence.

*2.2. Errors in distance measurements*

A variety of systematic errors affect the determination of stellar distances. The first source of errors could be from B and V magnitudes. This affects the distances of individual stars, but should not affect mean distance of a sample of stars, since the photometric errors should be randomly distributed. More important are the systematic errors that could be produced by the contamination by subgiant, giant and binary stars in our samples.

2.2.1. Subgiant contamination

We used the main-sequence color-absolute magnitude relation to derive the distances. For the distance determination, it is firstly assumed that all stars are unevolved. This is certainly not applicable to all stars in our survey, but without spectroscopy or photometric gravity indicators there is no *a priori* way to separate main-sequence dwarfs from evolved subgiant stars at the main-sequence turnoff. In our selection of subsamples of F and G−type stars ($3.5 \leq M_V \leq 6$ or $0.3 \leq$ B-V $\leq 0.9$), the contamination by giant stars should be small because at the above absolute magnitude and color intervals, distances and velocities implied for red giant stars are improbable. However contamination of main sequence ($4 \leq M_V \leq 6$) by subgiants with $2 \leq M_V \leq 4$ (with same color) could cause a serious underestimate of distance.

We have made a selection by color, by raising the blue limit of the color range (B-V$\leq 0.9$) beyond the region where subgiant stars are expected to contribute substantially and we can greatly reduce their contamination. Only drawback of this selection is that we loss the farthest distance and most of the halo stars. The Besançon model of population synthesis (Robin & Crézé 1986; Bienaymé et al. 1987) has been used to check the proportion of subgiants in our subsample of stars. If we suppose that there exists an intermediate-metallicity population of stars with scale height $h_z \sim 750$ pc and local density of 6% relative to the thin disk, then we would expect 30% of the intermediate subgiant population ($M_V \sim 4$) to have ($14 \leq$ V $\leq 16$) and the bulk to have V$\sim 17$. An old population I distribution ($h_z \sim 250$ pc) would have subgiants ($M_V \sim 4$) contributing mostly at V$\leq 14.5$. The number of possible disk and thick disk subgiants in our survey is minimal, since the total number of stars in the appropriate magnitude and color range is about a 10%.

For the halo, the distribution is fairly broad. We would expect the largest contribution from halo subgiants at $15.5 \leq$ V $\leq 19.5$. The worst contamination of the main sequence by spheroid giants probably occurs at $0.6 \leq$ B-V$\leq 0.8$ or $5.9 \leq M_V \leq 6.7$ (Sandage 1982, Bahcall et al. 1983). Recently, a spectroscopic observations have been performed on a sample of stars in a field of 9.6 square degrees in the galactic centre direction ($l \simeq 3°$, $b \simeq 45°$) by Perrin et al. (1995). The most striking result of these observations is the very low proportion of Population II giants in this direction, about 2 times less than the predictions of the current galactic models. Similar result has been found in other intermediate latitude fields by different authors (Friel 1987, Morrison 1993).

It is possible that some ranges of (V, $M_V$) are not seriously contaminated. For example, if in the range $4 \leq M_V \leq 6$, the luminosities of the subgiants are close to their main sequence values then the mis-identified evolved stars do not come from a much larger volume (Tritton & Morton 1984). Also at distances where the disk density dominates over the spheroid, subgiant or giant contamination should not be a problem.

2.2.2. Binary contamination

Another systematic error affecting the distance determination is the problem of unresolved binaries. Every star has been treated as if it were a single object, but this is certainly not the case for a significant portion of the sample "but unknown". Only binaries with components of approximately equal mass are problematical. In this case, the distance error will be 40%, while for slightly unequal mass stars, there is a concern for color errors in the primary (a 10% error in color results in 0.6 mag error in estimated absolute magnitude). The binary fraction of Population I stars is supposed to be relatively large-approximately 55%. The thick disk binary fraction is essentially unknown, but it is likely to be somewhere between that of the thin disk and the halo. In summary, systematic errors in the distance determination due to the presence of binaries are probably most severe for young disk stars. Clearly this problem would be solved with a systematic spectroscopic survey for the sample of stars at moderately high resolution to identify spectroscopic binaries. Recently, such a programme has been initiated by Soubiran (1994) for subsamples of stars in our surveys.

### 2.2.3. Photometric errors

The major factors contributing to distance errors are the photometric errors in the color $\sigma_{B-V}$ and magnitude $\sigma_V$. To roughly estimate the contribution from all of these, it is necessary to go back to the observable quantities, as some of these errors are correlated, e.g. $\sigma_{B-V}$ and $\sigma_{[Fe/H]}$ with $\sigma_{M_V}$. From our *in situ* surveys, we expect the mean errors in $\sigma_{B-V} = 0.1$ and $\sigma_V = 0.07$ for V = 11 to 18. From these values, our estimate of error to be about 20 % in distance seems to be realistic up to z = 3 kpc. However, we have estimated that for farthest distances (z $\geq$ 3 kpc), there may be an error in our distances as large as 30 %.

## 3. Tangential velocities

The cardinal components of the stellar space velocity (in km/s), U, V and W (where U is defined as positive in the direction of the Galactic anticentre, V is positive in the direction of Galactic rotation, and W is positive in the direction of NGP) were derived from proper motions, $\mu_l$ and $\mu_b$ (in arcsec year$^{-1}$), and distance d (in pc) for the three fields as follows (Murray 1983) :

$$\begin{pmatrix} U \\ V \\ W \end{pmatrix} = R_G \begin{pmatrix} 4.74\,d\,\mu_l \\ 4.74\,d\,\mu_b \\ V_r \end{pmatrix}$$

The matrix $R_G$ for each field is as follows :

$$GAC1,2 \quad \begin{pmatrix} 0.7497 & 0.0368 & -0.6608 \\ 0.0808 & 0.9859 & 0.1466 \\ 0.6569 & -0.1633 & 0.7381 \end{pmatrix}$$

$$GC \quad \begin{pmatrix} 0.5301 & -0.5049 & 0.6813 \\ 0.6674 & 0.7440 & 0.0320 \\ -0.5230 & 0.4377 & 0.7314 \end{pmatrix}$$

$$GAR \quad \begin{pmatrix} 0.8753 & -0.4746 & 0.0931 \\ 0.4043 & 0.6123 & -0.6794 \\ 0.2655 & 0.6323 & 0.7278 \end{pmatrix}$$

The unknown radial velocity contributions ($V_r$) to UVW are as follows and are neglected in the following study :

GAC1,2 : $U_{V_r} = 0.661\,V_r$; $V_{V_r} = 0.147\,V_r$; $W_{V_r} = 0.738\,V_r$

GC : $U_{V_r} = -0.681\,V_r$; $V_{V_r} = 0.032\,V_r$; $W_{V_r} = 0.731\,V_r$

GAR : $U_{V_r} = -0.093\,V_r$; $V_{V_r} = -0.680\,V_r$; $W_{V_r} = 0.728\,V_r$

It is clear from the above 3 matrices that we are measuring the velocities (U-W,V) in the direction of GAC1,2, (U+W,V) in the direction of GC and (U,V-W) in the direction of GAR field. We should point out here that the measured proper motions are directly converted to $U \pm W$ and $V$ velocities. For GC or GAC1,2 field :

$$\frac{U \pm W}{\sqrt{2}} \simeq 4.74\,d\,\mu_b \quad and \quad V \simeq 4.74\,d\,\mu_l$$

and we define :

$$\sigma^2_{U,W} = \frac{\sigma^2_U + \sigma^2_W}{2}$$

## 4. Space density

We are interested in deriving the structural parameters (e.g. scale height and scale length) of the thin and thick disk populations using our data sets at intermediate latitude. For this we have calculated the stellar space density as a function of both absolute magnitude and distance from the Galactic plane. The calculation is straightforward :

The space density of total stars (Population I + Intermediate Population + Population II) of the luminosity group ($M_1 < M < M_2$) falling into the partial volume $\triangle V_{1,2}$ is given by :

$$\rho(\bar{r}) = \rho(r_1, r_2) = N_{1,2}/\triangle V_{1,2}$$

$N_{1,2}$ being the total number of stars of the luminosity group in question, and partial volume $\triangle V_{1,2} = (\pi/180)^2(\square/3)(r_2^3 - r_1^3)$. $r_1$ and $r_2$ are the limiting distances; $\square$ : field size in square degrees; $\bar{r} = ((r_1^3 + r_2^3)/2)^{1/3}$ : centroïd distance of $\triangle V_{1,2}$.

### 4.1. Density laws

The space density distribution can be approximated by a double exponential :

$$\rho(R,z) = \rho(R_0) exp\left(-\frac{R-R_0}{h_R}\right) exp\left(-\frac{|z|}{h_z}\right) \quad (1)$$

$R_0 = 8.09 \pm 0.3$ kpc (Pont et al. 1994) is the solar distance from the galactic centre. $\rho(R_0)$ is the solar "neighborhood" normalisation. $R$ is the galactocentric distance projected upon the galactic plane and $z$ is the height above the galactic plane. $h_z$ and $h_R$ are the scale height and scale length, respectively.

The density can be expressed as a function of the distance d along the line of sight for a field of galactic coordinates $(l,b)$ with,

$$R = (R_0^2 + d^2 cos^2 b - 2R_0 d\,cosb\,cosl)^{1/2}$$

$$|z| = d\,sinb$$

We derived the logarithmic space density functions of the thin and thick disk stars in $3.5 \leq M_V \leq 5$ and $5 \leq M_V \leq 6$ absolute magnitude intervals. In deriving space density, we excluded data for all stars with distances outside the interval 150<z<2750 pc. These precautions are necessary to reduce the contamination of FGK dwarf sample by possible Population I and II giants, and to keep completeness of counts in all studied fields. The 150 pc limit then arises both because we have very few stars nearer than V≤11 and because of possible contamination by young disk stars. The outer distance limit of z≃2750 pc arises because the exponential fit is no longer valid beyond this distance due to completeness.

The observed density is fitted by the sum of two exponentials of the form shown in Equ. (1). A least-squares method is adopted to derive the best model in which the exponential distributions of the number density reproduce the observed starcounts. Bins of distance have been selected in order to have at least 50 stars per bin to reduce the Poisson error. The values of the scale lengths of thin and thick disk populations determined from the star count ratio (see §5.3) were fixed in the calculation. We find that varying these scale lengths, the value of scale height changes only slightly. To avoid these uncertainties in the determination of structural parameters, we have used the combination of three intermediate latitude fields (GAC1,2, GC & GAR) explained in the next subsection. We computed a number of exponential models iteratively until the quantity $\sum_{p=1}^{P} (N_p^{obs} - N_p^{model})^2$ is minimized on a domain of the major parameters. $p$ is the number of bins. It is evident that, in the least-squares solution, the bins are weighted nearly proportional to the number of stars located in them. We found that the least-squares solution for the intermediate component is not greatly changed. In figures 1 and 2, we present the density distribution of stars for two different absolute magnitude ranges as a function of distance above the galactic plane.

Tables 1 and 2 present the best fitted structural parameters for stars with $3.5 \leq M_V \leq 5$ and $5 \leq M_V \leq 6$ absolute magnitude intervals from the two data sets (GAC1,2 and GC).

By comparing table 1 and table 2, we find some differences on scales depending both on fields and absolute magnitude intervals. These differences may be due to the larger statistical errors in the GAC1,2 data set (area 7 sq. degs. and 5956 stars) compared to the GC data set (area 15 sq. degs. and 19696 stars). They may also be produced by errors in B-V color that is used to select the samples.

4.2. The density distribution $\rho(z)$ for stars with $3.5 \leq M_V \leq 5$ derived from the three intermediate latitude fields (GAC1,2, GC and GAR)

We have used the combination of 3 intermediate latitude fields (GAC1,2, GC and GAR) to derive the structural parameters of the two populations. To minimize the un-certainty in the determination of density laws, we built a more consistent model. We fit simultaneously data in each field with the same values of parameters (expect for the thin disk scale length that has not been fitted). For each disk, we have used the same local normalization of density independently of directions. In this case we may solve both parameters ($h_R$ & $h_z$) together. The observed density is fitted by the sum of two exponentials shown in figure 3. The same procedure (described in the previous section) was applied to derive the best fitted structural parameters. The results are shown in table 3.

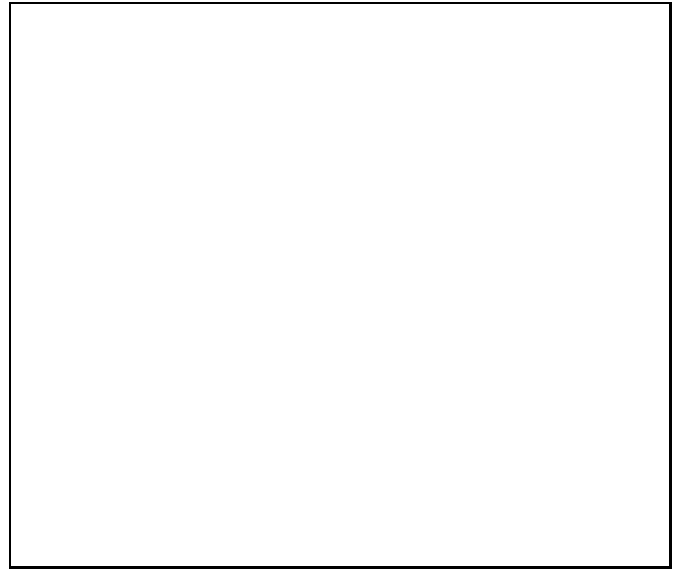

**Fig. 1.** The density distribution for stars with $3.5 \leq M_V \leq 5$ as a function of distance above the galactic plane. The continuum line represents the sum of two exponentials with scale heights given in table 1 and corresponding to the thin disk and thick disk

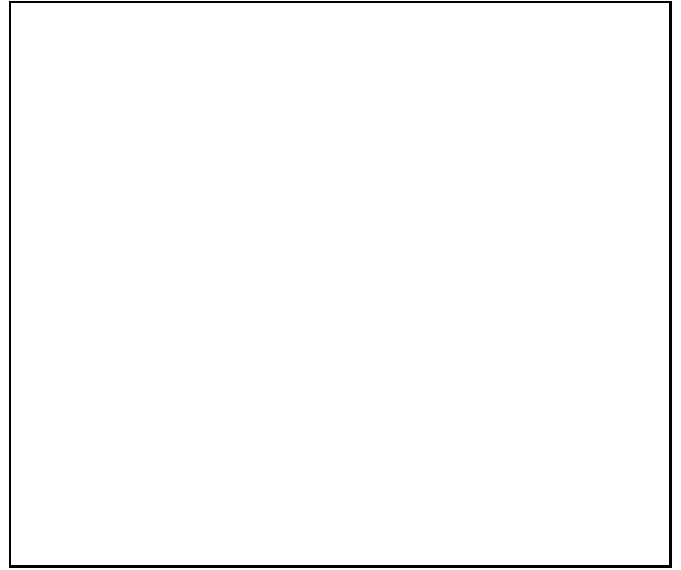

**Fig. 2.** The density distribution for stars with $5 \leq M_V \leq 6$ as a function of distance above the galactic plane. The continuum line represents the sum of two exponentials with scale heights given in table 2 and corresponding to the thin disk and thick disk

**Table 1.** The best fitted structural parameters of the thin and thick disk stars with $3.5 \leq M_V \leq 5$ derived from the two data sets (GAC1,2 and GC)

| Field | Thin Disk | Thick Disk | Thin Disk : Thick Disk |
|---|---|---|---|
| $3.5 \leq M_V \leq 5$ | $h_z$ (pc) | $h_z$ (pc) | density ratio |
| GAC1,2 | 336 | 807 | 100 : 7.3 |
| GC | 226 | 674 | 100 : 9.8 |

**Table 2.** The best fitted structural parameters of the thin and thick disk stars with $5 \leq M_V \leq 6$ derived from the two data sets (GAC1,2 and GC)

| Field | Thin Disk | Thick Disk | Thin Disk : Thick Disk |
|---|---|---|---|
| $5 \leq M_V \leq 6$ | $h_z$ (pc) | $h_z$ (pc) | density ratio |
| GAC1,2 | 267 | 849 | 100 : 8.6 |
| GC | 217 | 726 | 100 : 9.9 |

### 4.3. Discussion

The best value for the **scale height of thin disk is $h_z$ = 260±50 pc**. In contrast with the generally adopted value around 325 pc, we find a lower value for the scale height. This low value is in agreement with the scale height of 249 pc fitted by Kuijken & Gilmore (1989) from a K dwarf photometric parallax study in the direction of the South Galactic Pole. Haywood (1994, 1995) showed that the overall vertical density profile of the galactic disk is closed to an exponential with scale height $h_z \simeq 250$ pc. Ng et al. (1995) recent determination gives $h_z = 250$ pc for the scale height of thin disk based on a sample of stars toward the galactic centre.

The **thick disk characteristics are, $h_z = 760\pm50$ pc and local density = $7.4^{+2.5}_{-1.5}$ % relative to the thin disk**. Robin et al. (1995) recent determination gives $h_z$=760±50 pc, with a local density of 5.6±1.0 % relative to the thin disk. This last determination has been done with available B, V photometric data taken quite extensively from literature. They use a different approach to analyse data using a synthetic model reproducing observable quantities (magnitude and color counts). This method is expected to avoid systematic bias that can be encountered in inversing the process. Two points can explain the small differences between Robin et al. (1995) and present determinations. First, the density law used by Robin et al. (1995) is slightly different from our own. They use a modified exponential with a null derivative at z = 0, in order to conform to the potential. Our density law is strictly exponential thus the local density deduced from the counts at z>1 kpc is slightly overestimated in our case compared to Robin et al. (1995) result. Second, there is a slight correlation between scale height and density when determined using star counts (Reid & Majewski, 1993). However, Robin et al. (1995) used a larger magnitude intervals (from V = 12 to 22), thus raising the indetermination.

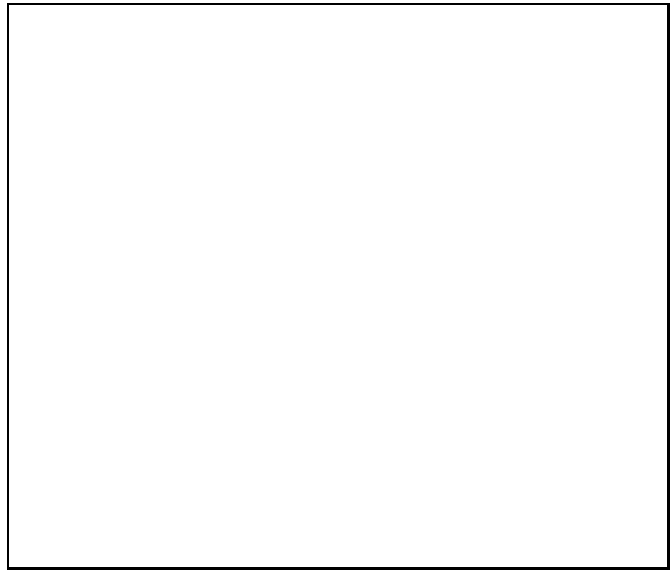

**Fig. 3.** The density distribution for stars ($3.5 \leq M_V \leq 5$) in three fields (GAC1,2, GC and GAR) as a function of distance above the galactic plane. The continuum line represents the sum of two exponentials with scale height 260 pc and 760 pc and corresponding to the thin disk and thick disk, respectively

**Table 3.** The best fitted structural parameters for the thin disk and thick disk stars with $3.5 \leq M_V \leq 5$ derived from the three data sets (GAC1,2, GC and GAR). The scale length of the thin disk was not well determined due to poor statistics of the data in the nearer distance bins (z<800 pc)

| Thin Disk | Thick Disk | | Thin Disk : Thick Disk | |
|---|---|---|---|---|
| $h_z$ (pc) | $h_z$ (pc) | $h_R$ (pc) | density ratio | |
| 260±50 | 760±50 | 3800±500 | 100 : | $7.4^{+2.5}_{-1.5}$ |

Our scale height determinations are mainly based on the shape of densities on figures 1-3, while the scale lengths depend on ratio of counts between opposite fields. Small relative photometric errors between fields will not introduce large change on the shape of figures, but will affect relative counts between fields and modify the scale length. This effect is larger when we select sample with small extent in $B - V$ (e.g. the values of the scale length of thick disk range between 2 and 3 kpc for the second sample ($5 \leq M_V \leq 6$), and 3.7±0.5 kpc from the first sample ($3.5 \leq M_V \leq 5$)). A conservative estimate of the errors consists to consider the maximal fluctuations of our various determinations for different samples and different fitting processes. We found that the scale length is not accurately measured but conversly the exact values have no large influence on the other measured quantities (scale height and local densities).

We deduce that the **scale length of the thick disk is $h_R = 3\pm1$ kpc**. Robin et al. (1995) determination gives 2.8±0.8 kpc, which is based partly on the same data. These results can be compared to recent determination given by Soubiran (1993ab), based on a sample of stars with proper motions towards the NGP : $h_{z,thin\ disk}$=280 pc, $h_{z,thick\ disk}$=700 pc. Von Hippel & Bothun (1993) gives $h_{z,thin\ disk}$ ∼290 pc, $h_{z,thick\ disk}$=860±90 pc based on a faint Strömgren photometric survey.

## 5. Kinematics of stellar populations : SEM (Stochastic-Estimation-Maximization) algorithm

The kinematical properties of each stellar population are related to their spatial distributions. Scale height, velocity dispersions and asymmetric drift are linked by the Boltzmann equation. The proportion of each population varies with the distance above the galactic plane and the selection of a stellar sample at a given distance allows to optimize the proportion of one population. Since kinematical data allow to improve this identification of populations, we have minimized bias from mutual contamination of each population at a given height by performing a kinematical separation of populations. For that purpose we have assumed that the kinematic of each population is well approximated by a maxwellian velocity distribution. In case of an isothermal population and separation of vertical and radial motions, this is particularly suited, since no kinematical gradient is expected, mean velocity dispersion and asymmetric drift remain constant with height above the plane. This allows to recognize one population at various heights and to measure its density along the line of sight.

To perform the kinematical separation, we have used a maximum likelihood method (SEM algorithm : Celeux & Diebolt 1986) in order to deconvolve the multivariate gaussian distributions and estimate the corresponding parameters. The aim of the SEM algorithm is to resolve the finite mixture density estimation problem under the maximum likelihood approach using a probabilistic teacher step. Full details can be found in a review paper published by Celeux & Diebolt (1986). Through SEM one can obtain the number of components of a gaussian mixture (without any assumption on this number), its mean values, dispersions and the percentage of each component with respect to the whole sample. This method has already been used by Gómez et al. (1990) and Soubiran (1993ab) to characterize the (U,V,W) parameters of the stellar populations.

The samples of stars in 2 fields (GAC1,2 and GC) have been devided in 6 or 7 bins of distance, and in each bin of distance a fit has been performed with a SEM algorithm to separate the 2-D gaussian distributions to identify the three components (thin disk, thick disk and halo) of the Galaxy. As can be seen in Ojha et al. (1994a) (tables 11 & 12) and Ojha et al. (1994b) (tables 10 & 11), the percentage ratio of 2 populations (thin disk/thick disk) varies with the distance above the galactic plane and therefore with the line of sight distance. The thin disk population has been identified as a mixture of components (probably young and old disks) and thus we observe a gradient in velocity dispersions as a function of z. The thick disk population (table 12 in Ojha et al. (1994a) and table 11 in Ojha et al. (1994b)) has been identified as a discrete and distinct component. We do not find any gradient either radial or vertical in the velocity ellipsoid of the thick disk population. This proves that the thick disk is an isothermal population. The halo population has been identified in the farthest distance bins (z>2 kpc) with large velocity dispersions and in a *prograde rotation*. We present here new results obtained from the comparison of data from different fields that had been studied separately.

### 5.1. Kinematics of the thin disk population

For the small distance intervals r<500 pc, the gaussian fit shows that the thin disk-like population was itself divided into two components. This may be due to the varying contribution of young disk and old disk populations in these distance intervals. We observe a continuous increase of the velocity dispersion with height. It can be explained by the fact that the thin disk has formed over about 10 Gyr, therefore can be considered as a sum of isothermal disks. The dynamical evolution of the orbits results in a vertical gradient of velocity dispersion, detected in this analysis.

Using our two data sets in the directions of GC and GAC1,2, we obtain a measure of the radial gradient of velocity dispersion ($\frac{\partial ln\sigma^2_{U,W}}{\partial R}$), which is found to be -0.18±0.03 kpc$^{-1}$ or (5.7±1 kpc$^{-1}$) for stars (in color range 0.3<B-V<0.9) on a base line of ~ 2 kpc (figure 4). We obtain also ($\frac{\partial ln\sigma^2_V}{\partial R}$)=-0.05±0.02 kpc$^{-1}$. These results are in agreement with Fux & Martinet (1994) ($\frac{\partial ln\sigma^2_U}{\partial R}$ = -0.16$^{+0.02}_{-0.03}$ kpc$^{-1}$). The local value of the velocity dispersion gradient ($\frac{\partial ln\sigma^2_U}{\partial R}$) may be determined using the stability criterion of the galactic disk introduced by Toomre (1964). Based on the Toomre's criterion, Mayor (1974) deduced a value of -0.2 kpc$^{-1}$ for $\frac{\partial ln\sigma^2_U}{\partial R}$ in contrast to the value of zero that one would have in the ellipsoidal hypothesis. This result has been confirmed by Oblak & Mayor (1987) from F, G and K type stars in the Gliese catalogue, who found values between -0.14 and -0.24 kpc$^{-1}$. Our value for ($\frac{\partial ln\sigma^2_{U,W}}{\partial R}$) is also in good agreement with the estimation of Vandervoort (1975) and Erickson (1975) obtained from the 3rd and 4th order moments of the local stellar velocity distribution (between -0.19 and -0.23 kpc$^{-1}$) based also on marginal stability assumption and with Lewis and Freeman (1989) kinematical scale length of 4.4±0.3 kpc, ($\frac{\partial ln\sigma^2_U}{\partial R}$ = -0.23±0.02 kpc$^{-1}$) determined using a large number of (about 600) distant old disk K giants.

### 5.2. Kinematics of the thick disk population

#### 5.2.1. Gaussian deconvolution

The SEM solutions of the thick disk derived from each field are presented in Ojha et al. (1994ab). The mean kinematical results of the thick disk obtained from the 2 fields (GAC1,2 and GC) are summarized in table 4.

Galactic radial gradients of the velocity dispersion ($\frac{\partial ln\sigma^2}{\partial R}$) of the thick disk population were never determined before. In our present study, the thick disk population appears to be isothermal, since the kinematical properties are nearly constant along the line of sight. For this reason, the velocity characteristics for this population are much better defined. We determine a small positive, null and negative gradients respectively in GC, GAR and GAC1,2 fields, that may be explained as a radial gradient of the kinematic of the thick disk. Mean values allow to determine local properties of the thick disk. From comparison of proper motion distributions in each field, the kinematical gradients can be estimated on a base line of 5 kpc : $\frac{\partial ln\sigma^2_s}{\partial R}$ range from -0.02 to -0.10 kpc$^{-1}$.

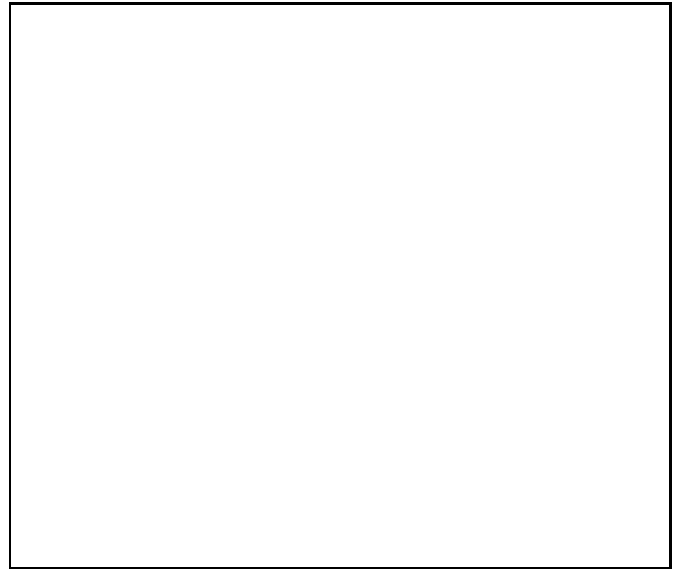

**Fig. 4.** Plot of $\sigma_{U,W}$ as a function of $d$ distances along the line of sight at color interval 0.3<B-V<0.9 mag for the two data sets : galactic anticentre field ($l = 167°$, $b = 47°$) and galactic centre field ($l = 3°$, $b = 47°$)

**Table 4.** Kinematics of the thick disk population (expressed in km/s) derived from SEM algorithm. Velocities are with respect to the Sun

| Anticentre (GAC1,2) | | Centre (GC) | |
|---|---|---|---|
| $<\frac{U-W}{\sqrt{2}}>$ | -5±6 | $<\frac{U+W}{\sqrt{2}}>$ | 4±3 |
| <V> | -57±4 | <V> | -49±3 |
| $\sigma_{U,W}$ | 64±4 | $\sigma_{U,W}$ | 66±3 |
| $\sigma_V$ | 60±3 | $\sigma_V$ | 56±2 |

By combining the kinematical results deduced from 4 fields (GAC1,2, GC, GAR and NGP (Soubiran 1993ab)), we have derived the velocity ellipsoid of the thick disk population. The mean kinematic parameters are summarized in table 5 and discussion is given below.

Soubiran (1993ab) and Kharchenko et al. (1994) have determined kinematical properties of the thick disk from proper motion surveys towards the NGP. Their results can be compared to our results since selection criteria are very similar. Applying our distance determination (see §2) and SEM to their samples, we obtain nearly the same values for dispersions and asymmetric drift. Then we obtain a mean $\sigma_U = 71\pm5$ km/s from measurements towards the NGPs and from our field (GAR). We mention that this value is consistent with the dynamical constraint of the asymmetric drift relation (see §6). $\sigma_V$ had been obtained in four fields (three different directions : GC, GAC1,2 and NGP) and is apparently constant : mean $\sigma_V = 57\pm4$ km/s.

It is difficult to extract $\sigma_W$ from our determinations of $\sigma_{U,W}$ and $\sigma_U$ using the relation $\sigma_{U,W}^2 = (\sigma_U^2 + \sigma_W^2)/2$, since errors from both determinations are added and $\sigma_W$ is the (small) difference of two large quantities. The relative error on $\sigma_W$ is then too large. We prefer to apply an indirect determination in order to estimate $\sigma_W$. The scale height of an isothermal population is related to its vertical velocity dispersion. The exact relation can be determined if the vertical potential is known. In fact this method has been frequently applied to determine the vertical potential from samples of disk stars with known vertical density distribution and velocity dispersions. Most recent results (Bienaymé et al. 1987; Kuijken & Gilmore 1989; Flynn & Fuchs 1994) have shown that the vertical potential can be explained by the visible mass (stellar and IMS) without advocating for a dark matter thin disk. Such determinations are based on disk stars not very distant from the galactic plane, where the vertical force changes linearly, and where the scale height of these populations is nearly proportional to their velocity dispersions. For thick disk stars considered in this paper, their distances range from 1 to 2 kpc, where the vertical force is nearly constant, and so the scale height of the population in this range of distance is nearly proportional to $\sigma_W^2$. So $\sigma_W$ is less dependent on the exact value of the potential. Taking the potential obtained by Bienaymé et al. (1987), Kuijken & Gilmore (1989) and Flynn & Fuchs (1994), we determine a value of $\sigma_W$ = 40 km/s. Assuming the potential of Oort (1965) with 30 percent of unseen matter in the disk, we would obtain $\sigma_W$ = 45 km/s.

The accuracy on measured dispersions depends on the number of stars in the sample. The samples towards the NGP are relatively small and result in larger errors (error bars given by Soubiran (1993ab) reflect directly the number of stars in her sample). Errors on Kharchenko et al. (1994) data must be similar (5 to 10 km/s). A second effect comes from the accuracy of proper motion that has been exactly determined for all these surveys, and range from 1 to 2 mas/year or 10 to 40 km/s for stars at distance of 1 to 2 kpc. From the observed proper motion distributions and with the help of the Besançon galactic model, we determine the ellipsoid of the velocity distribution corrected from errors on the proper motions (table 5).

*Remark* : All previous determinations have been done assuming that the main axis of the velocity ellipsoid remains parallel to the plane. The exact situation is not clear while it is sometimes claimed that the ellipsoid should point towards the plane in a direction beyond the galactic centre. If the ellipsoid is oriented in such a way, the velocity dispersion observed on the GC field would imply a smaller $\sigma_U$ in this direction, and a slightly larger in the GAC field. This implies a gradient of $\sigma_U$, with increasing values of $\sigma_U$ outwards. This apparently contradicts the fact that we observe a small negative gradient for the V-dispersions (that is not modified by an ellipsoid inclination), and since U and V-dispersions are kinematically linked.

**Table 5.** The mean kinematic parameters of thick disk (in km/s) derived from 4 fields (GAC1,2, GC, GAR & NGP). $\sigma_W$ is determined for the most probable vertical potential (see text). $V_{Lag}$ is with respect to the Sun

|  | $\sigma_U$ | $\sigma_V$ | $\sigma_W$ | $V_{Lag}$ |
|---|---|---|---|---|
| Thick disk | 67±4 | 51±3 | 40 | -53±10 |

It should be noted that since two years the accuracy of the kinematical measurements of the thick disk population has greatly improved and the results are well in agreement with each others (see table 6). It remains a controversy about a possible vertical gradient claimed by Majewski (1992) but not seen in data from Soubiran (1993ab), Ojha et al (1994ab) and others. The determination of the gradient is of great consequence on the scenario of formation for this population (see §7). The controversy seems to come from the way populations are distinguished from each others. The SEM algorithm (used by Soubiran 1993ab and Ojha et al. 1994ab) allows to well separate the thick disk and to evaluate its kinematics within a good accuracy in distance bins where it is preponderant.

5.2.2. Multivariate discriminant analysis

Because the kinematics and metallicity of thick disk population is supposed to differ from the thin disk and halo, it may be possible to pick out thick disk stars on the basis of the kinematical differences between the three populations. We therefore try here to find new constraints on thick disk population using samples at intermediate latitude which include photometry and proper motions. We have used multivariate discriminant analysis (MDA) to qualify the thick disk using observations in multidimensional space (V, B-V, U-B, $\mu_l$ & $\mu_b$).

A MDA is used to search for the best discriminant axes to project the data in multi-dimensional space to seek the optimal separation between different populations. The distribution of stars along the discriminant axis (combination of the observed axis) shows the clearest separation between the thin disk, thick disk and halo. We have applied

**Table 6.** Determinations of the velocity ellipsoid of thick disk population (in km/s)

| Author(s) | Thick disk | | | |
|---|---|---|---|---|
| | $\sigma_U$ | $\sigma_V$ | $\sigma_W$ | $V_{Lag}$ |
| Wyse & Gilmore 1986 | 80 | 60 | 60 | 100 |
| Carney & Latham 1986, 1989 | – | – | – | 30 |
| Norris 1986,1987ab | – | – | 35 | 20 |
| Sandage & Fouts 1987 | 75 | 35 | 42 | 31 |
| Ratnatunga & Freeman 1989 | 77±16 | 54 | – | – |
| Morrison et al. 1990 | 55 | 40 | – | 35±10 |
| Majewski 1992* | 35-130 | 35-70 | – | 21-120 |
| Soubiran 1993ab | 56±11 | 43±6 | – | 41±16 |
| Kharchenko et al. 1994 | 67±1 | 54±1 | – | 63±1 |
| Bartasiute 1994 | 64±5 | 49±3 | 42±3 | 39±5 |
| Beers & Larsen 1994 | 63±7 | 42±4 | 38±4 | 20 |
| Layden 1994 | 59±12 | 43±10 | 42±10 | 34±10 |

* The author advocates a vertical kinematical gradient of the intermediate population stars

this method to our data sets (GC and GAC1,2), using simulations from the Besançon model of population synthesis to separate the thick disk among other populations and to investigate the circular velocity of this population. The advantage of this method is that we can extract the kinematic parameters of stellar populations without estimating the stellar distances i.e. by using a model of population synthesis.

Using model predictions, we select a subsample of stars (where the thick disk stars are in majority), chosen as B-V≤0.8 and 14≤V≤15.5. We used the model simulations to find the best discriminant axes where to project the data in order to separate the thick disk population from the thin disk and halo. This was done using a multivariate discriminant analysis under the MIDAS[1] environment. Model simulations have been made assuming different circular velocities of thick disk. The characteristics of each tested model are shown in Ojha et al. (1994b) (table 8). To avoid too large Poisson noise in the Monto-Carlo simulations, we computed at least 10 simulations of 100 square degrees for each of the models tested in our analysis. The first discriminant axis for the circular velocity of 180 km/s of thick disk in the direction of GC is given by :

$$x = 0.024(B-V)+0.139(U-B)-0.079V-0.310\mu_l-0.069\mu_b$$

It should be noted that the discriminant axis varies when we change the circular velocity in model simulations. The resulting discriminant axis is dominated by the proper motion along the rotation ($\mu_l$) and by the U-B color index

[1] MIDAS is an acronym for München Image Processing and Data Analysis System developed by European Southern Observatory.

due to metallicity differences between the thin disk, thick disk and halo.

To quantitatively estimate the adequacy of the models with various circular velocities, we applied a $\chi^2$ test to compare the distribution of the sample on the discriminant axis with a set of model predicted distributions assuming different circular velocities of thick disk. The $\chi^2$ is given by the following formula :

$$\chi^2 = \sum_{i=1}^{n} \frac{(b_i - a_i)^2}{a_i}$$

Where n is the number of bins and $a$ & $b$ are the number of counts in each bin in the model and observed data sets, respectively. The resulting $\chi^2$ distribution may be approximated by a normal distribution given by the expression (Wilson & Hilferty 1931) :

$$\{(\frac{\chi^2}{n-1})^{1/3} + \frac{2}{9(n-1)} - 1\}(\frac{9(n-1)}{2})^{1/2}$$

The above expression is approximately normally distributed around zero mean with unit variance (Kendall & Start 1969).

Table 7 and figure 5 show the values of the probability (in sigmas) of each model to come from the same distribution as the observed sample. The most probable value for the lag or asymmetric drift of thick disk comes out to be -55±10 km/s (see figure 5), which shows that the circular velocity of thick disk is of the order of 180±10 km/s (where $V_{LSR}$ = 229 km/s and $V_\odot$ = 6.3 km/s in the model). We notice that the model predictions are at 3 sigmas of the data. This is due to the fact that the statistics of the errors in the data is not a Poisson statistics, because of possible systematic errors in the photometry and astrometry.

**Table 7.** $\chi^2$ test for models with different circular velocities for the thick disk. Lag or asymmetric drift is with respect to the Sun, where $V_{LSR}$=229 km/s and $V\odot$ = 6.3 km/s in the Besançon model

| $V_{cir}$ (km/s) | $V_{Lag}$ (km/s) | $\chi^2$ (in sigmas) | |
|---|---|---|---|
| | | GC | GAC1,2 |
| 150 | -85.3 | 8.7 | 6.8 |
| 165 | -70.3 | 6.8 | 5.7 |
| 175 | -60.3 | 4.1 | 3.8 |
| 180 | -55.3 | 4.3 | 3.4 |
| 185 | -50.3 | 4.9 | 3.5 |
| 190 | -45.3 | 5.5 | 4.0 |
| 215 | -20.3 | 6.7 | 5.1 |

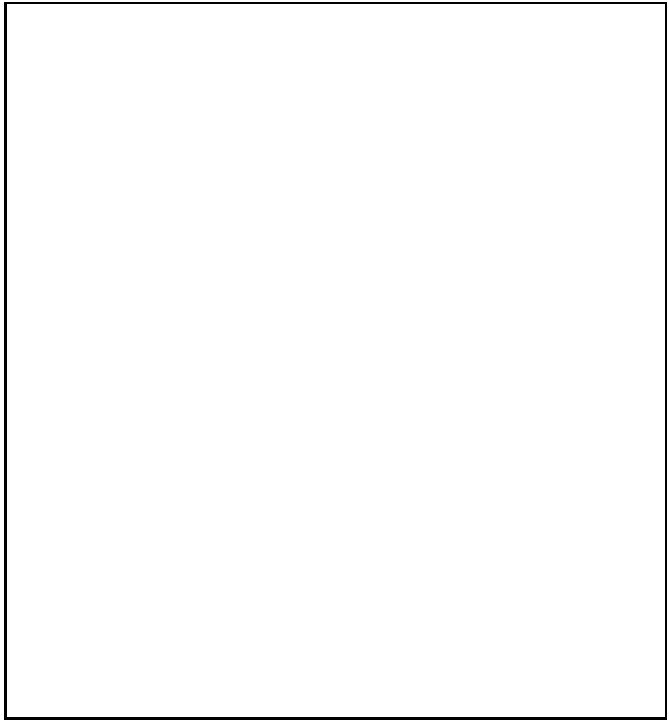

**Fig. 5.** $\chi^2$ test for models with different circular velocities for the thick disk population for the two fields (GC and GAC1,2). $\chi^2$ is given in number of sigmas

We obtain a unique value in both directions, showing that no radial gradient seems to occur on a base of 3 kpc around the Sun. The distribution over the discriminant axis of observed stars (solid points) towards GAC1,2 ($l$ = 167°, $b$ = 47°) with best model predictions (with three populations) is shown in figure 6.

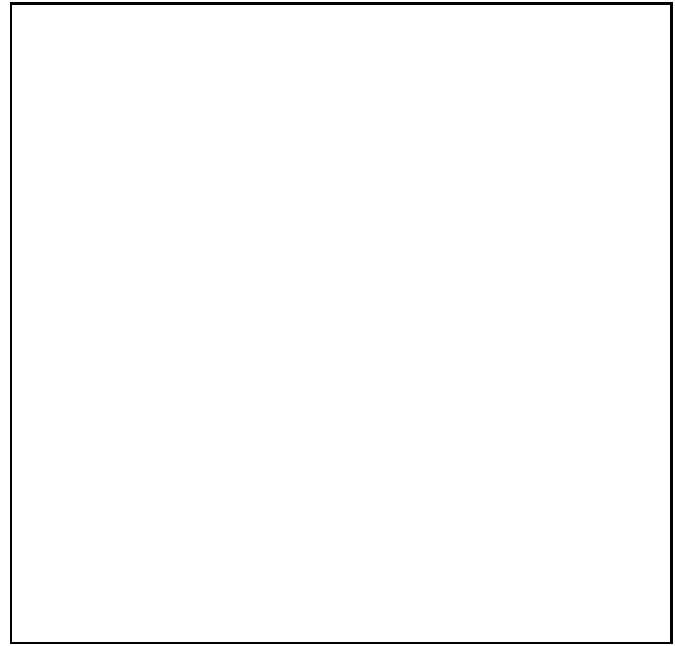

**Fig. 6.** Distribution over the discriminant axis of observed stars (solid points, with $\pm\sqrt{N}$ Poisson statistics) towards galactic anticentre direction; best model predictions (solid full thick line) and model predicted stars according to their populations (thick line : thin disk, full thick line : thick disk and thin line : halo)

The resulting asymmetric drift of thick disk ($V_{Lag}$ = -55±10 km/s) is in good agreement with the determination made using the gaussian deconvolution of velocity distribution (i.e. $V_{Lag}$ = -53±10 km/s, see §5.2.1). The MDA approach we use here to separate the thick disk from other populations, allows to avoid two biases. First, by using a model of population synthesis, which adopts the same selections in the model and data, and second, by identifying the thick disk from other populations on a physical basis, by its metallicity and kinematics and using at the same time all 5 observables.

### 5.3. Scale lengths from kinematically identified populations

In section 4, we have determined the scale length and scale height of the populations from computing the density law along the line of sight in different absolute magnitude (or color) intervals. The kinematical separation of populations using SEM algorithm also gives an estimation of the proportions and densities of each population along the line of sight distance. Figures 7 and 8 show the number of F and G type stars of the thin and thick disk respectively, as a function of line of sight distance ($d_{los}$).

Using the density function of Equ. (1) the star count ratio between the two fields GC and GAC1,2 (where $|R - R_0| \simeq |z|$), assuming that the luminosity function is the same, can be written as :

$$A_{GC}(m)/A_{GAC1,2}(m) = exp(+2 * |R - R_0|/h_R) = exp(\sqrt{2}\, d_{los}/h_R)$$

or

$$h_R = \frac{\sqrt{2}\, d_{los}}{log(A_{GC}(m)/A_{GAC1,2}(m))}$$

By comparing the star count ratio between the two data sets in each distance bin, we obtain the scale length of thin disk is $h_R = 2.6 \pm 0.6$ kpc and for the thick disk $h_R = 3.6 \pm 0.5$ kpc. It appears from the test with simulated data that SEM results were not very stable when two populations have the same proportions. This limit the accuracy of density determination in some distance bins of transitions. However the results obtained on scale length is compatible with the results obtained directly from star counts in §4 (table 3), comforting the separation of population made from the SEM method.

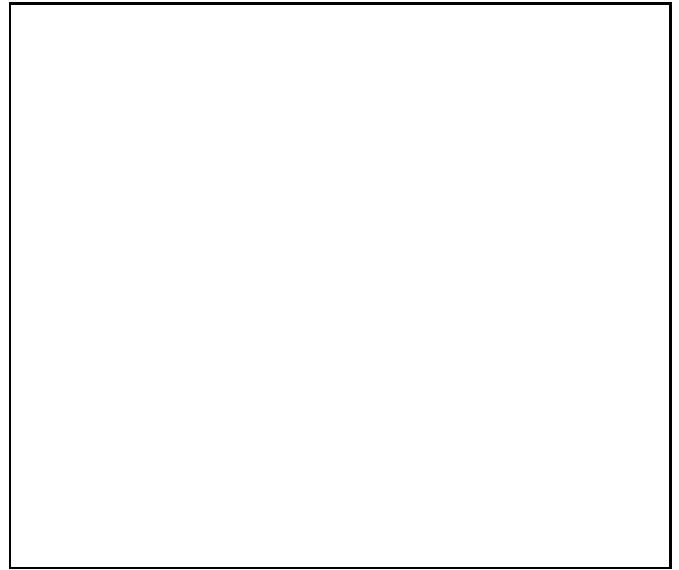

**Fig. 8.** The observed number of thick disk stars as a function of $d$ distances obtained from SEM algorithm from the two data sets (GAC1,2 and GC) in 0.3<B-V<0.9 color interval

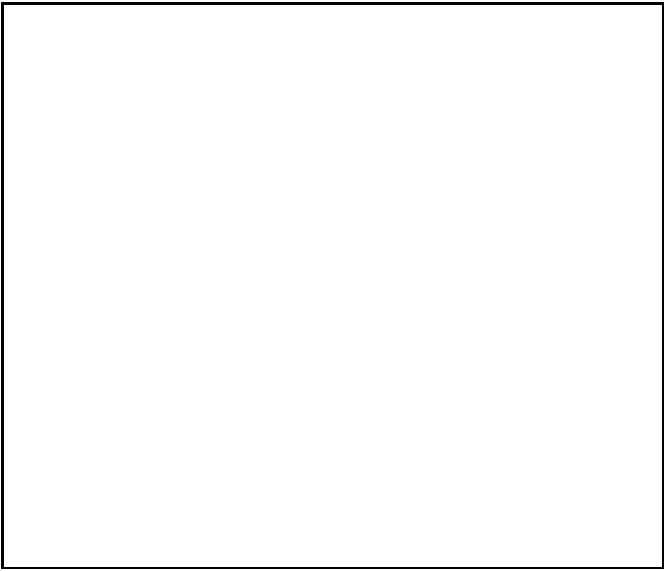

**Fig. 7.** The observed number of thin disk stars as a function of $d$ distances obtained from SEM algorithm from the two data sets (GAC1,2 and GC) in $0.3 \leq$B-V$\leq 0.9$ color interval

## 6. Solar motion and asymmetric drift relation

### 6.1. Solar motion

Velocity components of the solar motion are usually deduced directly from the mean motion of nearby stars relative to the Sun. The tangential component $V_\odot$ of the solar motion in a circular rotating coordinate frame is based on the extrapolation of the asymmetric drift relation (<V>, $\sigma_U^2$), where <V> is the mean of the V velocity components of a stellar sample, $\sigma_U^2$ is the variance. Here, we measure the solar motion from distant sample of stars (1 to 3 kpc) and compare it to the local determinations.

Since our samples of stars are non local, we deduce the asymmetric drift relation from a general form (see Binney & Tremaine, 1987) and we obtain the expression :

$$-V_{lag} = -V_c(R) + V_{rot}(R) = \frac{R\sigma_U^2}{2V_c}\left[\frac{\partial ln\rho}{\partial R} + \frac{\partial ln\sigma_U^2}{\partial R}\right.$$
$$\left. + \frac{1}{R}(1 - \frac{\sigma_V^2}{\sigma_U^2}) + \frac{1}{\sigma_U^2}\frac{\partial \rho <UW>}{\rho \partial z}\right] \quad (2)$$

or shortly

$$-V_c(R) + V_{rot}(R) = \frac{\sigma_U^2}{D} \quad (2.1)$$

under the following assumptions that the Galaxy is stationary and axisymmetrical, and $V_{lag} << V_c$. $V_c(R)$ is the circular velocity curve and $V_{rot}(R)$ is the mean rotational velocity of stars with radial velocity dispersion $\sigma_U$.

The mean apparent velocity of a stellar sample relative to the Sun is given by :

$$<V>_{obs} = V_{rot}(R) - V_c(R_\odot) - V_\odot \quad (3)$$

combining the equations (2 & 3), we obtain :

$$<V>_{obs} \sim \frac{\partial V_c(R)}{\partial R}(R - R_\odot) - V_\odot + \frac{\sigma_U^2}{D}$$

In figure 9, we have plotted the $(<V>, \sigma^2_{U,W})$ points (with $\sigma^2_{U,W} = (\sigma^2_U + \sigma^2_W)/2$) obtained from the observational data sets in two directions (GC and GAC1,2) for F and G-type stars (0.3<B-V<0.9). We find that, for $\sigma_{U,W}$ < 19 km/s, the $(<V>, \sigma^2_{U,W})$ points differ significantly from the straight line (see also Mayor 1974). In two papers, Mayor (1970, 1972) have shown that density waves or local perturbations disturb essentially the stellar populations with low velocity dispersions. Then discarding these subpopulations having a low velocity dispersions and within the limits of statistical uncertainties, the extrapolation of $\sigma = 0$ gives a mean $V_\odot = 6.6 \pm 1.6$ km/s. Joined with the mean values of the U and W components, we obtain the solar motion : $U_\odot$ = -1.0±6.0 km/s, $V_\odot$ = 6.6±1.6 km/s and $W_\odot$ = 6.4±6.0 km/s.

The derived standard solar motion is typical for F and G-type disk dwarfs. These values can be compared to those of Mayor (1974) : (-10.3±1.0, 6.3±0.9, 5.9±0.4) km/s and Oblak (1983) : (-8.2±1.8, 5.0±0.7, 5.5±0.4) km/s, determined from the local sample of stars in a radius of 200 pc around the Sun. There is no significant differences between the values of solar motion derived from local and distant samples of stars, implying no visible mean motion of local stars relatively to distant one on the meridional plane.

### 6.2. Galactic rotation curve

The value of D depends on quantities that are equal in both opposite fields (GC and GAC1,2). The contribution of the gradient of the circular velocity curve has an opposite sign in the two fields, and comparison of them gives a direct measure of $\nabla V_c(R)$. On figure 9, the relation $(<V>, \sigma^2)$ has nearly the same value for both fields, indicating a null gradient for the velocity curve. We determine $\nabla V_c(R) = +0.7 \pm 1$ km/s based on a range of distances of ±1.5 kpc and conclude that the rotation curve is flat around the solar radius. Our analysis is in agreement with various determinations of the outer rotation curve (Fich & Tremaine, 1991) based on more distant tracers.

### 6.3. Asymmetric drift

The gradient of the velocity curve being null, the expression of the asymmetric drift can be written in a simpler form:

$<V>_{obs} = -V_\odot + \frac{\sigma^2_U}{D}$

Simplified expressions for D are frequently found and are generally based on assumptions without any strong theoretical support. A careful analysis can be found in Fux & Martinet (1994). Current assumptions concern the last term in the bracket of Equ. (2) that is frequently replaced by its value for a spherical or a plane parallel potential, but not by its more probable value which may be intermediate. For $z = 0$, this last term can be written (Fux & Martinet 1994) as $\frac{\lambda(R)}{R}(1 - \frac{\sigma^2_W}{\sigma^2_U})$ with $0 \leq \lambda \leq 1$.

Exact value of $\lambda$ depends on unknown quantities like the flattening of the dark halo. In the following discussion we will choose $\lambda = 0.7$, according to Fux & Martinet (1994).

The other assumption concerns the kinematical scale lengths that is currently put equal to the density scale length. This is partly based on Lewis and Freeman (1989) kinematical observations. Their kinematical gradient seems to be confirmed by few existing measurements including those presented in this paper. However the value they adopt for the density scale length is now very questionable and is much larger than the most recent direct (non-kinematical) determinations. Then we cannot assess equality of density and kinematical scale length for our own Galaxy.

If $z \neq 0$, the vertical derivative of density is not zero. Since we are analysing samples of stars far outside the mid-plane of the Galaxy, a new term appears inside the bracket of Equ. (2). For an exponential disk of the form $\rho(z) = exp(-(z/h)^n)$, we obtain :

$$\frac{2V_c V_{lag}}{R\sigma^2_U} = \frac{2(A-B)}{D} = -\frac{1}{H_\rho} - \frac{1}{H_{\sigma^2_U}} + \frac{1}{R}(1 - \frac{\sigma^2_V}{\sigma^2_U})$$
$$+ \frac{\lambda(R)}{R}(1 - n(\frac{z}{h})^n)(1 - \frac{\sigma^2_W}{\sigma^2_U})$$

where A and B are Oort's constants and $H_\rho$, $H_{\sigma^2_U}$ are density and kinematical scale lengths. This relation shows that for an isothermal population, the asymmetric drift does not change with the distance above the galactic plane if and only if the potential is cylindrical ($\lambda = 0$). We can estimate for typical values ($\lambda = 0.7$, n=1) that the lag of a population with $\sigma_U = 30$ km/s will vary by only 1 km/s on one scale height, and for $\sigma_U = 70$ km/s by 5 km/s.

#### 6.3.1. Thick disk

We have informations for all dominant terms present in the asymmetric drift relation for the thick disk population. We have measured the scale length directly from star counts, the asymmetric drift, $\sigma_V$ in all fields, $\sigma_{U,W}$ for GC and GAC1,2 fields, and we get $\sigma_U$ from independent proper motion surveys. So we can determine one of these quantities from the others, and try to determine the self-consistency of observed quantities.

Since the scale height of thick disk is 760 pc and the distances of observed stars range between 1 to 2 kpc above the galactic plane where the $K_z$ force is nearly constant and the density is exponential, we get $n = 1$ and $z/h \sim 1.5 - 3$ in the last term of the asymmetric drift relation. We apply the asymmetric drift relation in order to obtain an estimate of $\sigma_U$ by replacing all the other known quantities. Velocity gradients are null for the thick disk population. Substituting $V_{Lag}$ = -53 km/s (with respect to the Sun), $V_c$=220 km/s, $h_R$=2.8 kpc, R=8.09 kpc and $\lambda = 0.7$ in Equ. (2), we obtain $\sigma_U$= 72–82 km/s, in agreement with the value given in table 5 (67 ± 4 km/s). Conversely, we may determine the "density" scale length by replacing $\sigma_U$ by the value given in table 5. We find from Equ. (2), the scale length of thick disk to be 2–2.8 kpc.

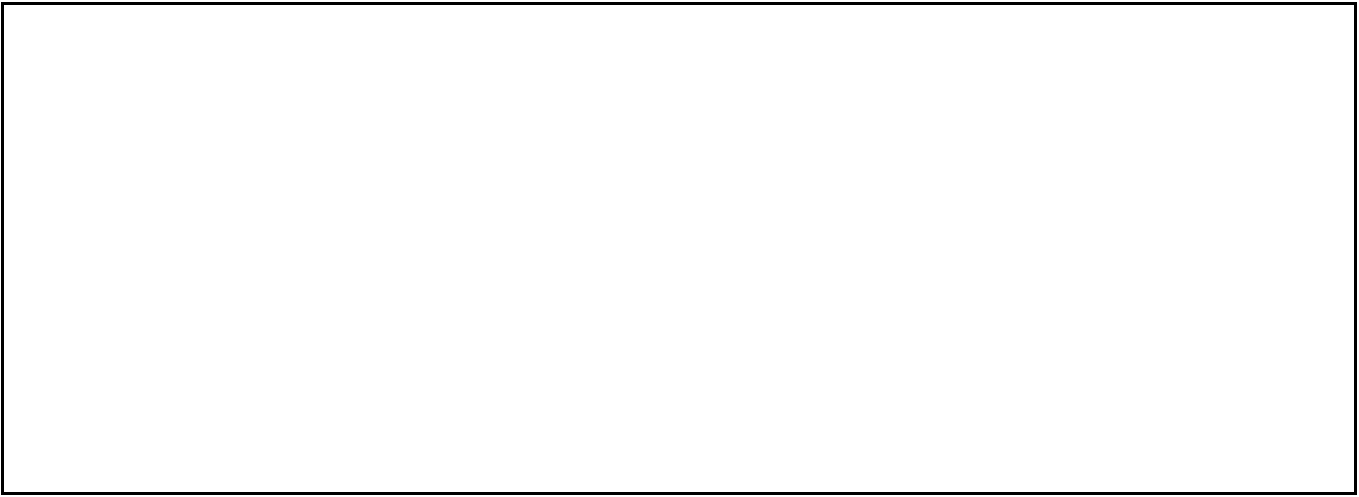

**Fig. 9.** Asymmetric drift <V> *versus* $\sigma^2_{U,W}$ for F and G-type stars in 2 fields (GAC1,2 and GC). The regression line gives a tangential motion for the Sun of $V_\odot$ = 6.6±1.6 km/s

Substitutions of all observed quantities in the asymmetric drift equation show the consistency of our kinematical data with scale length determined by star counts.

### 6.3.2. Thin disk

In the same way, we have tried to determine the scale length of the thin disks from the asymmetric drift relation. The thin disk is certainly made of a set of continuous disks with varying scale heights, lengths and kinematics. This could be the reason why we do not succeed to extract from Equ. (2) a simple relation to predict the relation found on figure 9. For that purpose it would be necessary to know exactly the proportion of each isothermal component at each height. It appears also that the exact value of the index $n$ giving the shape of the vertical potential is determinant. Moreover, it does not seem that the relation should be necessary linear. Such uncertainties lead to very large range of possible scale length (positive or negative!).

## 7. Discussion

In the following subsection, we discuss various determinations of the disk's radial scale length. In the next subsection, we discuss some of our results in terms of the history of the galactic disk and thick disk populations, its present state and the formation scenarios of the thick disk.

### 7.1. The thin disk's radial scale length

There have been several determinations of the radial scale length in the thin disk of the Milky Way either from photometric or kinematic approach.

The many different determinations based on photometry or star counts give possible values in a large range (Kent et al. 1991 give a recent review, see also Robin et al. 1992). To add to this apparent confusion, many published density scale lengths are deduced from kinematic data using the asymmetric drift relation, but generally the expressions used for the asymmetric drift are simplified without any strong theoretical support. For example, it is frequently assumed that the kinematic and density scale lengths are equal. Only in a recent contribution (Fux & Martinet 1994) the term including the shape (spherical, cylindrical...) of the potential is adjusted.

Here, we have presented new direct measurements of density and kinematic gradients. The resulting density scale length of 2.3±0.6 kpc determined by our data is the lower limit of currently accepted values. It has to be compared to some recent works : Robin et al. 1992, $h_R$ = 2.5 kpc, from star counts towards the galactic anticentre, Fux & Martinet 1994, $h_R$ = 2.5 kpc, based on a rigorous analysis of the asymmetric drift relation.

The kinematic gradient presented here $\frac{\partial ln \sigma^2_{U,W}}{\partial R}$ = -0.18±0.03 kpc$^{-1}$ is consistent with a few other direct determinations. Density and kinematic scale lengths we obtain are consistent in respect to the asymmetric drift relation.

### 7.2. The formation of the thick disk

Possible thick disk formation scenarios have been summarized by Gilmore et al. (1989) and Majewski (1993). One of these scenarios is that the formation of the thick disk occurs due to kinematically heating of thin disk. This heating may have been violent, perhaps from the accretion of a satellite galaxy (Carney et al. 1989, Quinn et al. 1993). So the merging of one or more smaller galaxies with our Galaxy is a possible explanation for the formation of a thick disk. In such a scenario, the thick disk could be either the direct relic of the merged galaxy, dissolved and smoothly distributed in our Galaxy, or an old thin disk formed before the merger event(s) and 'puffed up' by the gravitational perturbations of the merging galaxy. In case of mergers between disks and satellites, both the radial and vertical structures of the disk are altered. The radial heating and re-arrangement of material by angular momentum transport results in an asymmetric drift (Wyse 1994). The vertical heating increases the disk scale height by about a factor of two. Direct evidence for an on-going merger event has been recently discovered towards the galactic bulge by Ibata et al. (1994).

Recently Quinn et al. (1993) (hereafter QHF) have presented a model of thick disk, which is based on a scenario of satellite accretion by a spiral galaxy which may produce a thick disk if the event occurs at the beginning of the life of the thin disk. According to this model the satellite accretions do gradually thicken disks by depositing ordered kinetic energy into random motions of disk stars. The abruptness of accretion origin implies a kinematical, chemical and age disjointedness from the thin disk.

Of course the N-body simulations of QHF is only illustrative, but our measurements reproduce most of features of their numerical simulations. Asymmetric drift ($\sim$ 50 km/s) and scale height (760 pc) of the thick disk are in good agreement with the QHF model. Our measurement of a vertical velocity dispersion of 40 km/s for the thick disk is also well in agreement with the QHF accretion model. Also the velocity dispersion gradient of the QHF's thick disk is much smaller than for the disk, and is nearly zero in the outer part. This corresponds to the null gradient observed. Scale lengths of thin disk and thick disk are also in agreement if we are in the outer part of the QHF model. This estimate is suggestive, however, that the thick disk model, which has a vertical velocity dispersion of $\sim$ 45 km/s, could be formed from a thin disk with vertical dispersion of $\sim$ 20 km/s by accretion of a satellite of about 10% of the mass of the disk.

Our measurement of the scale height of thick disk (760 pc) is also in perfect agreement with QHF model. However, our data do not allow to observe the correlation of scale height with the galactocentric radius as QHF have predicted in their N-body simulations.

Finally the local surface density of the thick disk is about 18% of thin disk, this gives a crude estimate of the total mass ratio of thick disk and thin disk stars. This implies that the merging occurs quite early in the Galaxy life when 5-10% of the disk was formed. If the star formation rate has been nearly constant (Haywood et al. 1995) at the epoch of formation, this indicate that the merge occurs 2 to 4 billions years after the formation of disk component (of course this primordial-disk has been transformed in thick disk).

The main feature we determine for the thick disk is its isothermality and its clear separation from thin disk and halo. This component is clearly simple and not a mixture of sub-components as for the thin disk and there is no gradient in the kinematic neither vertically nor radially. Similar feature has been obtained by Gilmore et al. (1995) and Robin et al. (1995), who find a null vertical gradient in metallicity : this last result can be understood as a consequence of the isothermality of the thick disk.

Concerning alternative explanations, we mention Burkert et al. (1992) who build a 1-D (vertical) model of collapse of our Galaxy and formation of galactic disk. Their self-regulated chemical and dynamical evolution of an initially hot, gaseous protodisk leads to the formation of a thick disk. This is the only "top-down" scenario of formation for the thick disk where some discontinuity between thin and thick disk is predicted with an acceptable agreement with our observations. However a more accurate chemico-dynamical models (Samland, 1994), with a 2-D resolution show a more continuous age, metallicity and kinematic gradient between the various disk components.

To sum up, the results emerging from the present study of the correlations between photometry and kinematics give a mounting evidence in favour of merging process of satellite galaxies with the disk of our Galaxy for the formation of the thick disk.

## 8. Conclusion

In this paper we have used star counts and kinematical data to constrain the galactic structure parameters. We have also shown that a proper statistical analysis of the data in the 5-dimensional space (V, B-V, U-B, $\mu_l$, $\mu_b$) in comparison with the model of population synthesis allows one to constrain the physical model parameters.

We obtain a value of -0.18±0.03 kpc$^{-1}$ for the galactic radial gradient of velocity dispersion ($\frac{\partial ln\sigma_{U,W}^2}{\partial R}$) for the thin disk population. Determined from moderately distant stars, a new measurement of the solar motion is obtained. We found no systematic motion of the LSR relative to distant stars on the galactic meridional plane. The rotational velocity curve is found flat in the solar neighborhood. Our results confirm that the thin disk has a relatively short scale length of 2.3±0.6 kpc and scale height of 260±50 pc. The thick disk population is distinct from other populations based on their kinematical and spatial distributions. The data constrain the asymmetric drift of the intermediate population, which is found to be 53±10 km/s with respect to the Sun. No radial or vertical gradient is found in the rotational velocity and velocity dispersions of the thick disk population. We therefore conclude that the formation of the thick disk did not occur as a smoothly con-tinuous transitional phase between formation of the halo and formation of the thin disk during the collapse of the Galaxy (Eggen, Lynden-Bell & Sandage 1962). Rather, it support the model of a scenario of satellite accretion by our Galaxy which may produce a thick disk if the event occurs at the beginning of the life of the thin disk (Quinn et al. 1993). The most probable value of scale height for the thick disk component is determined to be $h_z \simeq 760 \pm 50$ pc with a local density of $A_{thick} = 7.4^{+2.5}_{-1.5}$ % relative to the thin disk. The ratio of the number of thick disk stars in our galactic centre region to that in anticentre region yields $h_R \sim 3 \pm 1$ kpc for the scale length of thick disk. These values are in perfect agreement with the recent determination given by Robin et al. (1995) based on the analysis of a large set of available photometric catalogues with accurate photometry.

*Acknowledgements.* This research work was partially supported by the Indo-French Centre for the Promotion of Advanced Research (IFCPAR) / Centre Franco-Indien Pour la Promotion de la Recherche Avancée (CFIPRA), New Delhi (India). We especially thank referee Dr. Gerry Gilmore for his comments. We also thank Caroline Soubiran and Elena Schilbach for letting us use their data.

**References**

Armandroff T., Zinn R., 1988, AJ 96, 92
Baade W., 1944, ApJ 100, 137
Baade W., 1958a, in Stellar Populations, ed. D.J.K. O'Connell (Amsterdam, North Holland), p. 3
Baade W., 1958b, in Stellar Populations, ed. D.J.K. O'Connell (Amsterdam, North Holland), p. 303
Bahcall J.N., Soneira R.M., Morton D.C., Tritton K.P., 1983, ApJ 272, 627
Bartašiūtė S., 1994, Baltic Astronomy, vol. 3, p. 16
Beers T.C., Sommer-Larsen J., 1995, ApJS 96, 175
Bienaymé O., Robin A.C., Crézé M., 1987, A&A 180, 94
Bienaymé O., Robin A.C., Crézé M., 1990, IAU symp 144, The interstellar disk-halo connection in Galaxies - poster proceedings, p. 5
Bienaymé O., Mohan V., Crézé M., Considère S., Robin A.C., 1992, A&A 253, 389
Binney J., Tremaine S., 1987, Galactic Dynamics, page 198, Princeton University press
Blaauw A., 1965, In Stars and Stellar systems, Vol. 5, In Galactic structure, eds. A. Blaauw & M. Schmidt (Chicago : University of Chicago press), p. 435
Burkert A., Truran J.W., Hensler G., 1992, ApJ 391, 651
Butler D., Kinman T.D., Kraft R.P., 1979, AJ 84, 993
Butler D., Kemper E., Kraft R.P., Suntzeff N.B., 1982, AJ 87, 353
Carney B.W., 1979, ApJ 133, 211
Carney B.W., Latham D.W., 1986, AJ 92, 60
Carney B.W., Latham D.W., Laird J.B., 1989, AJ 97, 423
Carney B.W., Latham D.W., Laird J.B., 1990, AJ 99, 572
Celeux G., Diebolt J., 1986, Rev. de Statistiques Appliquées 34, 35
da Costa G.S., Ortolani S., Mould J., 1982, ApJ 257, 633
Eggen O.J., Lynden-Bell D., Sandage A., 1962, ApJ 136, 748
Erickson R.R., 1975, ApJ 195, 343
Fenkart R., 1988, A&AS 76, 469
Fich M., Tremaine S., 1991, ARA&A 29, 409
Flynn C., Fuchs B., 1994, MNRAS 270, 471
Friel E.D., 1987, AJ 93, 1388
Fux R., Martinet L., 1994, A&A 287, L21
Gilmore G., Reid N., 1983, MNRAS 202, 1025
Gilmore G., Wyse R.F.G., 1985, AJ 90, 2015
Gilmore G., Wyse R.F.G., 1986, Nature 322, 806
Gilmore G., Wyse R.F.G., Kuijken K., 1989, ARA&A 27, 555
Gilmore G., Wyse R.F.G., Jones J.B., 1995, AJ 109, 1095
Gómez A.E., Delhaye J., Grenier S., et al., 1990, A&A 236, 95
Haywood M., 1994, Ph.D. thesis, Paris Observatory, France
Haywood M., Robin A.C., Crézé M., 1995, A&A (accepted)
Ibata R.A., Gilmore G., Irwin M., 1994, Nature 370, 194
Janes K.A., 1979, ApJS 39, 135
Jennens P.A., 1975, MNRAS 172, 695
Kendall M.G., Stuart A., 1969, The Advanced Theory of Statistics, Vol. 1, chap. 16
Kent S.M., Dame T.M., Fazio G., 1991, ApJ 378, 131
Kharchenko N., Schilbach E., Scholz R.-D., 1994, Astron. Nachr. 315, 291
King I.R., 1971, AJ 83, 377
Kuijken K., Gilmore G., 1989, MNRAS 239, (605, 651)
Laird J.B., Carney B.W., Latham D.W., 1988, AJ 95, 1843
Layden A., 1994 (private communication)
Lewis J.R., Freeman K.C., 1989, AJ 97, 139
Lindblad B., 1925, Ark. Mat. Astr. Fys., 19A, Nos. 21, 27, 35; 19B, No.7
Lindblad B., 1927, MNRAS 87, 553
Lindblad B., 1959, in Encyclopedia of Physics, Vol. 53, ed. S. Fliigge (Berlin, Springer-Verlag), p. 21
Majewski S.R., 1992, ApJS 78, 87
Majewski S.R., 1993, ARA&A 31, 575
Mayor M., 1970, A&A 6, 60
Mayor M., 1972, A&A 18, 97
Mayor M., 1974, A&A 32, 321
McClure R., Crawford D., 1971, AJ 76, 31
Morgan W.W., 1959, AJ 64, 432
Morrison H.L., Flynn C., Freeman K.C., 1990, AJ 100, 1191
Mould J.R., 1982, ARA&A 20, 91
Murray C.A., 1983, Vectorial Astronomy, p. 271, Adam Hilger Ltd. Bristol
Ng Y.K., Bertelli G., Bressan A., Chiosi C., Lub J., 1995, A&A 295, 655
Norris J., 1986, ApJS 61, 667
Norris J., 1987a, ApJ 314, L39
Norris J., 1987b, AJ 93, 616
Norris J.E., Green E.M., 1989, ApJ 337, 272
Oblak E., 1983, Ph.D. thesis, Besançon Observatory, France
Oblak E., Mayor M., 1987, Evolution of Galaxies, X IAU European meeting, ed. J. Palous, Publ. Astron. Inst. Czech. Acad. Sci. 69, 263
Ojha D.K., Bienaymé O., Robin A.C., et al., 1994a, A&A 284, 810 (GAC1)
Ojha D.K., Bienaymé O., Robin A.C., et al., 1994b, A&A 290, 771 (GC)
Ojha D.K., Bienaymé O., Mohan V., Haywood M., Robin A.C., 1995, A&A (in preparation) (GAR or GAC2)
Oort J.H., 1922, BAIN, No. 23
Oort J.H., 1926, Groningen Obs. Pub., No. 40

Oort J.H., 1927, BAIN 3, 275
Oort J.H., 1965, In Galactic Structure, ed. A. Blaauw & M. Schmidt, p. 455, Chicago: University of Chicago Press
Perrin M.-N., Friel E.D., Bienaymé O., Cayrel R., Barbuy B., Boulon J., 1995, A&A 298, 107
Pont F., Mayor M., Burki G., 1994, A&A 285, 415
Quinn P.J., Hernquist L., Fullagar D.P., 1993, ApJ 403, 74
Ratnatunga K.U., Freeman K.C., 1989, ApJ 339, 126
Reid N., 1993, MNRAS 265, 785
Reid N., Majewski S.R., 1993, ApJ 409, 635
Robin A.C., Crézé M., 1986, A&A 157, 71
Robin A.C., Crézé M., Mohan V., 1992, A&A 265, 32
Robin A.C., Haywood M., Crézé M., Ojha D.K., Bienaymé O, 1995, A&A (accepted)
Rodgers A.W., Harding P., Sadler E., 1981, ApJ 244, 912
Samland M., 1994, Ph.D. thesis, Keil University, Germany
Sandage A., 1982, ApJ 252, 553
Sandage A., 1986, ARA&A 24, 421
Sandage A., 1987, AJ 93, 610
Sandage A., Fouts G., 1987, AJ 92, 74
Shapley H., 1918, ApJ 48, 154
Soubiran C., 1992, Ph.D. thesis, Paris Observatory, France
Soubiran C., 1993a, A&A 274, 181
Soubiran C., 1993b, IAU Symp. No. 161, Astronomy from wide field, imaging Potsdam (Germany) ed(s). H.T. MacGillivray et al., p. 435
Soubiran C., 1994, IAU Symp. No. 164, Stellar Populations, The Hauge (Netherlands), ed(s). P.C. van der Kruit & G. Gilmore, p. 386
Strömberg G., 1924, ApJ 59, 228
Toomre A., 1964, ApJ 139, 1217
Tritton K.P., Morton D.C., 1984, MNRAS 209, 429
van den Bergh S., 1975, ARA&A 13, 217
Vandervoort P.O., 1975, ApJ 195, 333
von Hippel T., Bothun G.D., 1993, ApJ 407, 115
Wilson E.B., Hilferty M.M., 1931, The Distribution of $\chi^2$, Proc. Nat. Acad. Sci., U.S.A. 17, 684
Wyse R.F.G., Gilmore G., 1986, AJ 91, 855
Wyse R.F.G., 1994, special session on Digitizing the Sky, summer AAS meeting, Minnesota
Yoshii Y., Ishida K., Stobie R.S., 1987, AJ 92, 323
Yoss K.M., Hartkopf W.I., 1979, AJ 84, 1293
Zinn R., 1980, ApJ 241, 602
Zinn R., 1985, ApJ 293, 424